\documentclass{iopart}

\usepackage{graphicx}

\usepackage{moreverb}
\usepackage{url}

\begin{document}

\title[Data analysis and graphing]{Data analysis and graphing in an
  introductory physics laboratory: spreadsheet versus statistics
  suite}

\author{Primo\v{z} Peterlin}

\address{University of Ljubljana, Faculty of Medicine, Institute of
  Biophysics,\\
  Lipi\v{c}eva 2, SI-1000 Ljubljana, Slovenia}

\ead{primoz.peterlin@mf.uni-lj.si}

\begin{abstract}
  Two methods of data analysis are compared: spreadsheet software and
  a statistics software suite.  Their use is compared analyzing data
  collected in three selected experiments taken from an introductory
  physics laboratory, which include a linear dependence, a non-linear
  dependence, and a histogram.  The merits of each method are
  compared.
\end{abstract}

\submitto{\EJP}

\section{Introduction}
\label{sec:intro}

An experiment is not completed when the experimental data are
collected.  Usually, the data require some processing, an analysis,
and possibly some sort of visualization.  The three most commonly
encountered approaches to data analysis and graphing are
\begin{itemize}
\item manual method using grid paper
\item general purpose spreadsheet software
\item statistics software suite
\end{itemize}

The manual method involving grid paper and a pencil often still seems
to be the one favoured by the teachers.  Indeed it has a certain
pedagogical merit, for it does not hide anything in black boxes: the
student has to work through all the steps him- or herself.  It is also
the preferred option in the environments where students cannot be
expected to own a personal computer.  However, an increasing
proportion of students does have access to a computer, either on their
own initiative or as part of an organized initiative (\emph{e.g.,}
\cite{Zucker:2008}), and they find the method tedious, painstaking
and old-fashioned.  In all honesty, their teachers do not use this
method in their own research work.

General purpose spreadsheet software like Excel (Microsoft), Quattro
Pro (formerly Borland, Inc. now Corel, Inc.), or the freely available
OpenOffice Calc (formerly Sun Microsystems, Inc., now Oracle
Corporation) contain most functions required for data analysis and
graphing.  They are commonly available and the students are generally
well-versed in their use.  However, spreadsheets have their drawbacks.
They are difficult to debug, and they often require an educated
operator in order to produce a visually satisfying graph without
visual distractions.

Due to volume academic licenses, statistical software suites like SPSS
(formerly SPSS,Inc., now IBM), SAS (SAS Institute, Inc.), Stata
(StataCorp, Inc.), Statistica (StatSoft, Inc.), Minitab (Minitab,
Inc.), or S-PLUS (TIBCO Software, Inc.) have also become a viable
option.  They usually allow both a spreadsheet-like access and
programming using a scripting language, and generally produce an
excellent graphical output.

Finally, one also needs to mention the fourth group: specialized
software for scientific graphing such as Origin (formerly MicroCal,
Inc., now OriginLab), SigmaPlot (formerly Jandel Scientific, now
Aspire Software International), IGOR Pro (Wavemetrics, Inc.), or Prism
(GraphPad Software).  This is usually the preferred option in a
research laboratory, but with a price of around 1000~EUR per license,
it is usually considered too expensive for a classroom or student
laboratory use.

In this paper, we illustrate two methods, one involving a spreadsheet
and another one a statistics software suite, on a set of three
problems taken from the laboratory course prepared for first year
students of medicine, dental medicine and veterinary medicine
\cite{Bozic:2003}.  Two of them involve bivariate data (illustrating a
linear and a non-linear dependence) and one involves univariate data
(a histogram of radioactive decay).  As typical examples of each
class, two programs were selected: Microsoft Excel (version 12,
included in Microsoft Office 2007) and R (version 2.10.1).  Even
though one is a commercial software product and the other is a free
one, we consider them the most popular representatives of their
respective groups, which justifies their comparison.  Furthermore, the
features presented in R work unchanged in S-Plus, the commercial
implementation of S, while the spreadsheet examples work, except where
noted, besides in Excel also in the freely available OpenOffice Calc.


Microsoft first introduced Excel for Apple Macintosh in 1985, while
the version for Microsoft Windows followed two years later.  It
started to outsell its then main competitor, Lotus 1-2-3, as early as
1988, and the introduction of Microsoft Windows 3.0 in 1990 cemented
its position as the leading spreadsheet product with a graphical user
interface (GUI).  Microsoft Excel is now available for Microsoft
Windows and the MacOS X environments.

R \cite{Ihaka:1996,R:Manual} is both a language and an environment
for data manipulation, statistical computing and scientific graphing.
R is founded on the S language and environment \cite{Becker:1984}
which was developed at Bell Laboratories (formerly AT\&T, now Lucent
Technologies) by John Chambers, Richard Becker and coworkers.  S
strove to provide an interactive environment for statistics, data
manipulation and graphics.  Throughout the years, S evolved into a
powerful object-oriented language \cite{Becker:1988,Chambers:1998}.
Nowadays, two descendants exist which build on its legacy: S-Plus, a
commercial package developed by Tibco Software, Inc., and the freely
available GNU~R.  Its primary source is the Comprehensive R Archive
Network (CRAN), \url{http://cran.r-project.org/}.

In the rest of the paper, we first compare Excel and R on three cases
taken from a introductory physics laboratory: a linear and a
non-linear regression and a histogram.  We then discuss the positive
and the negative aspects of using either approach, and finally present
the main conclusions.

\section{Comparison}

\subsection{Case 1: Linear dependence}

Let us start with a simple example, in which the linear dependence
between the concentration and the conductivity of a dilute electrolyte
is examined.  At 5 different concentrations of electrolyte, the
student takes measurement of the resistance of a beaker filled with
electrolyte to the mark, with a pair of electrodes immersed and
connected into a Wheatstone bridge.  Using one known value for
conductivity, the student calculates the rest of conductivities from
the resistances, plots the points into a scatterplot, fits the best
line through the data points, and, measuring its resistance, finally
determines the unknown concentration of an additional sample.  

\begin{figure}
  \centering
  \includegraphics[width=0.6\linewidth]{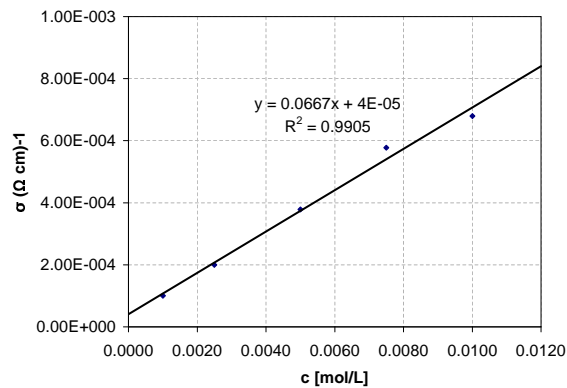}
  \caption{Using the ``Add trendline'' feature, linear dependence
    between the electrolyte concentration and its conductivity in a
    dilute electrolyte is easily determined with a spreadsheet.}
  \label{fig:lin-reg}
\end{figure}

The spreadsheet solution is straightforward: data points can be
plotted as a scatterplot, then with a right-hand click on any data
point, ``Add trendline'' is selected (figure~\ref{fig:lin-reg}).  In
order to learn the slope and the intercept of the straight line, one
can choose ``Display equation on chart'' in the ``Options'' tab.
Alternatively, one can find the slope, the intercept and the
correlation coefficient through built-in functions \texttt{SLOPE()},
\texttt{INTERCEPT()}, and \texttt{CORREL()}. More statistical
parameters can be obtained through a matrix function
\texttt{LINEST()}.  Individual cells from within the result of a
matrix function can be extracted by embedding \texttt{LINEST()} inside
a \texttt{INDEX()} function.  A fully worked example is given in the
on-line Supplementary Material.

\begin{figure}
  \centering
  \includegraphics[angle=270,width=0.6\linewidth]{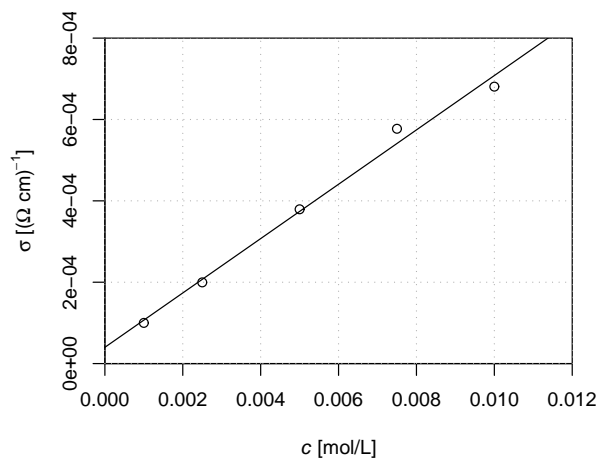}
  \caption{Linear dependence between the electrolyte concentration and
    its conductivity in a dilute electrolyte.}
  \label{fig:electrolyte}
\end{figure}

In R, the task can be achieved with a simple script, shown in Appendix
\ref{sec:appa}.  The resulting graph is shown in
figure~\ref{fig:electrolyte}.  The example shows a few features of R
which we want to comment on:
\begin{itemize}
\item Since we only have five data points, the data were entered
  directly into the program rather than being read from an external
  data file.  The \texttt{c()} function is used to concatenate data
  into a vector.
\item R encourages programming with vectors.  In the expression
  \texttt{k/resistance}, each element of the resulting vector is
  computed as a reciprocal value of the element of the original
  vector, multiplied by \texttt{k}.
\item Inspired by the \TeX\ typesetting system, R offers a capable
  method of entering mathematical expressions into graph labels using
  the \texttt{expression()} function \cite{Murrell:2000}.
\item The \texttt{plot()} function is the generic function for
  plotting objects in R; here we used it to produce a scatterplot.
\item The \texttt{lm()} function is used to fit any of several linear
  models \cite{Chambers:1991b}; we used it to perform a simple
  bivariate regression.  R is not overly talkative, and \texttt{lm()}
  does not produce any output on screen, it merely creates the object
  \texttt{cond.fit}, which we can later manipulate at will.  In the
  example we used \texttt{abline(cond.fit)} to plot the regression
  line atop the data points.  If we want to print the coefficients, we
  can use \texttt{coef(cond.fit)}.
\end{itemize}

In a simple case like linear dependence, spreadsheet offers a somewhat
simpler solution.  R, however, gives more control over its graphical
output and generally produces a visually more pleasing output.

\subsection{Case 2: Nonlinear dependence}

\begin{figure}
  \centering
  \includegraphics[scale=0.7]{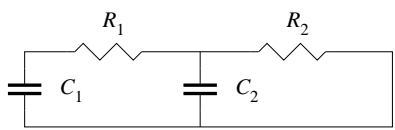}
  \caption{A circuit scheme with two capacitors.  In the experimental
    setup, $C_1 = C_2$ and $R_2 \gg R_1$.  $C_2$ is discharged at the
    beginning of the experiment.  The voltage $U$ on $C_1$ is recorded
    as a function of time $t$.}
  \label{fig:scheme}
\end{figure}

Not all dependencies are linear.  Sometimes, they can be linearized by
an appropriate transformation (\emph{e.g.,} logarithmic), which
reduces the task of finding the optimal fitted curve back to the
linear case.  With a computer at hand, however, this is not necessary.
In this example, we examine the time dependence of voltage in a
circuit with two capacitors (figure~\ref{fig:scheme}).  The student
first charges the capacitor $C_1$ and then monitors the voltage as
$C_1$ discharges.  The voltage $U(t)$ can be written as a sum of two
exponentials:
\begin{equation} 
  U(t) = A\, \mathrm{e}^{-t/\tau_1} + B\, \mathrm{e}^{-t/\tau_2} \; .
  \label{eq:two-exp}
\end{equation}
The task is to determine the two amplitudes, $A$ and $B$, and both
relaxation times, $\tau_1$ and $\tau_2$.

Solving the problem with a spreadsheet, one first learns that the
otherwise convenient ``Add trendline'' feature cannot be applied to
this case---while it offers a few functions beyond linear (polynomial,
logarithmic, exponential, power and running average), a sum of several
exponential functions is not among them.  Instead, we present two
other solutions in Excel.  The first one emulates the manual graphical
method, while the second one uses the built-in ``Solver'' tool.

\begin{figure}
  \centering
  \includegraphics[width=0.6\linewidth]{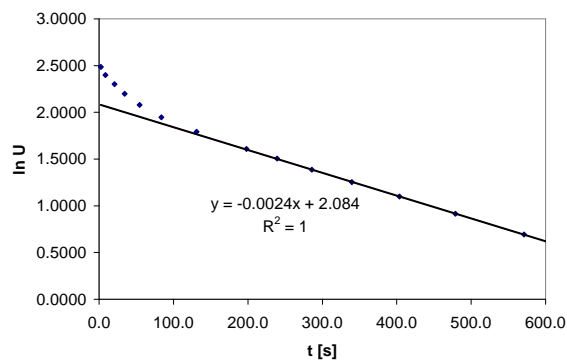}
  \caption{Logarithm of the dimensionless voltage $u =
    U/1\;\mathrm{V}$ on the capacitor $C_1$
    (figure~\protect\ref{fig:scheme}) as a function of time $t$; after
    $\approx 200$~s the fast component dies away and only the slow
    component remains, yielding $B$ and $\tau_2$.}
  \label{fig:2exp-log-1}
\end{figure}

The first solution (Supplementary Material, worksheet ``2exp log'')
takes into account the properties of the exponential function.  After
a certain period---this threshold value can be most easily obtained
graphically (figure~\ref{fig:2exp-log-1}) by plotting $\ln\,U$
(Supplementary Material, \texttt{supplement1.xls}, worksheet ``2exp
log'', column C) vs.\ $t$---the fast component, $A\,
\mathrm{e}^{-t/\tau_1}$, dies away, and only the slow component
remains:
\begin{equation} 
  U(t) \approx B\, \mathrm{e}^{-t/\tau_2} \; .
  \label{eq:slow-comp}
\end{equation}
Taking logarithm of (\ref{eq:slow-comp}) one obtains $\ln\,B$ as the
intercept (D13) and $-1/\tau_2$ as the slope (D12).  $\tau_2 =
410\;\textrm{s}$ and $B = 8.0\;\textrm{V}$ are computed in D16 and
D17, respectively.  Once the slow component is determined, one can
compute the fast component (column F) by subtracting the slow
component from the whole:
\begin{equation}
  \label{eq:fast-comp}
  U_h(t) =   U(t) - B\, \mathrm{e}^{-t/\tau_2} \; .
\end{equation}
From (\ref{eq:two-exp}) one can see that $U_h(t) = A\,
\mathrm{e}^{-t/\tau_1}$, therefore by taking the logarithm of
(\ref{eq:fast-comp}), one obtains $\ln\,A$ as the intercept (H13) and
$-1/\tau_1$ as the slope (H12), yielding $A = 3.7\;\textrm{V}$ and
$\tau_1 = 43\;\textrm{s}$.  This solution retraces the same steps one
would take if equipped only with a grid paper and a pocket calculator;
spreadsheet only makes it slightly more convenient.

While the linearized solution certainly possess certain pedagogical
merits, one needs to realize that the transformation needed to
linearize the data inevitably distorts the experimental error and
alters the relationship between the $x$- and $y$-values, yielding an
incorrect final result.  Instead, one can use the built-in ``Solver''
tool (the ``Solver'' included in the recent (3.1.1) release of
OpenOffice Calc is restricted to linear models and thus inappropriate
for the task) to maximize the coefficient of determination,
\begin{equation}
  R^2 = 1 - \frac{\sum_i(y - y_\mathrm{fit})^2}{\sum_i(y -
    \overline{y})^2} \; .
  \label{eq:coef-det}
\end{equation}
The spreadsheet solution is given in the Supplementary Material
(\texttt{supplement1.xls}, worksheet ``2exp nonlin'').  As before,
column A contains the independent variable ($t$) and column B the
dependent variable ($U$).  Column C contains the model curve
(\ref{eq:two-exp}), with the cells J3\ldots J6 containing the
coefficients $A$, $B$, $\tau_1$ and $\tau_2$, respectively.  Column D
contains the squared difference between the experimental values of the
dependent variable and the model curve, and column E the squared
difference between the individual values of the dependent variable and
their average value, the average value itself being saved in the cell
J7.  Finally, the cell J10 contains the coefficient of determination
(\ref{eq:coef-det}).

We fill the cells J3\ldots J6 with the initial guesses for the
parameters and start the Solver tool (Tools~$\rightarrow$~Solver).  We
want to maximize the coefficient of determination \cite{Brown:2001},
therefore we set J10 as our target cell, and set the ``Equal to''
option to ``Max''.  We enter the range J3:J6 into ``By changing
cells'' option and start the optimizer by clicking ``Solve''.  After
the solver has finished, we are asked whether we want the optimized
values of parameters to be written into the given cell range J3:J6.

\begin{figure}
  \centering
  \includegraphics[angle=270,width=0.6\linewidth]{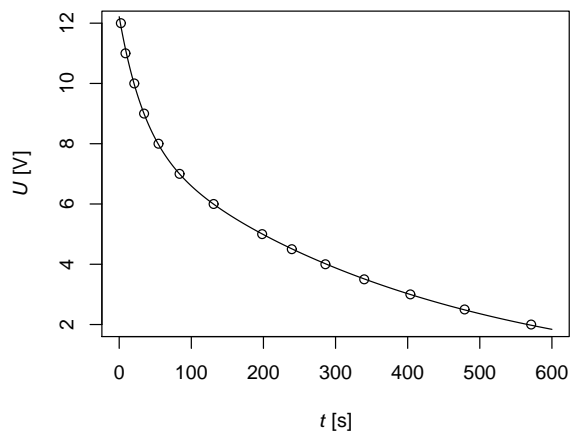}
  \caption{The voltage $U$ on the capacitor $C_1$
    (figure~\protect\ref{fig:scheme}) as a function of time $t$;
    measured data ($\circ$) and a fitted curve.}
  \label{fig:capacitor}
\end{figure}

The solution in R is shown in Appendix \ref{sec:appb}.  It assumes
that the data data is saved in a CSV format:
\begin{verbatim}
"t [s]";"U [V]"
2,3;12,0
8,7;11,0
...
\end{verbatim}
The resulting graph is shown in figure~\ref{fig:capacitor}.  A few
features of R used in the example need some comment:
\begin{itemize}
\item \texttt{read.table()} returns the object of a class
  \texttt{data.frame}; we may visualize it as a table.  Individual
  columns in the table are addressed as {\ttfamily{\itshape
      table}\/\${\itshape column}}.  Column names are constructed from
  the table header, with all illegal characters being replaced by
  dots.  For easier manipulation, two vectors, \texttt{t} and
  \texttt{U}, were created from the appropriate columns of the table.
\item The actual nonlinear fitting is performed by the \texttt{nls()}
  function, which returns an object of the \texttt{nls} class.  Apart
  from the formula, we also supplied the initial guesses for the
  fitting parameters.  The values of the coefficients can be extracted
  by \texttt{coef(U.nls)}, yielding $A = 4.0\;\mathrm{V}$, $B =
  8.2\;\mathrm{V}$, $\tau_1 = 35\;\mathrm{s}$, and $\tau_2 =
  403\;\mathrm{s}$.  The command \texttt{summary(U.nls)} produces an
  even more informative report.
\item An object of an \texttt{nls} class also provides methods for the
  \texttt{predict()} function, which we used to plot a piecewise
  linear curve, which appears as a smooth curve due to the small step
  used.
\end{itemize}

In this case, the solution in R does not appear any more complex than
the linear case---the considerably more difficult mathematics of the
non-linear regression are hidden from the user.  On the other hand, a
spreadsheet like Excel has no direct support for non-linear
regression, and subsequently the user has to work through some of the
necessary steps manually.

\subsection{Case 3: Histogram}

Our third example examines the statistics of radioactive decay.  Using
a Geiger counter, the student is required to record the number of
decays detected in a period of time (in our case, 10~s), and repeat
the measurement 100 times.  After subtracting the base activity, the
student computes the mean value $\overline{x}$, the standard deviation
$\sigma_x$, and plots a histogram.  

\begin{figure}
  \centering
  \includegraphics[width=0.6\linewidth]{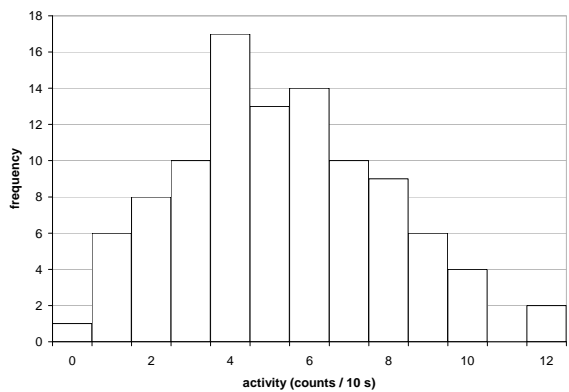}
  \caption{Lacking the support for histograms, one needs to resort to
    bar charts to plot a histogram of radioactive decay with a
    spreadsheet.}
  \label{fig:histogram}
\end{figure}

In a histogram, one plots the frequencies of the cases falling into
each of several categories, where the categories are usually specified
as non-overlapping intervals of the variable in question.  In a
spreadsheet, the \texttt{FREQUENCY()} function can be used to compute
the frequencies.  This is a matrix functions, which takes two ranges
of cells as input---the data array and the bins array---and
returns as output an array of the same length as the bins array.
While calculating the frequencies is supported, neither Excel nor
OpenOffice Calc provide a direct support for histogram plotting.
Since the bins were chosen to be of the same size in our example, we
can emulate them with bar charts (figure~\ref{fig:histogram}).  While
one can eliminate the gaps between bars in the chart setting, label
positions reveal the fact that we are dealing with a bar chart rather
than a histogram.

\begin{figure}
  \centering
  \includegraphics[angle=270,width=0.6\linewidth]{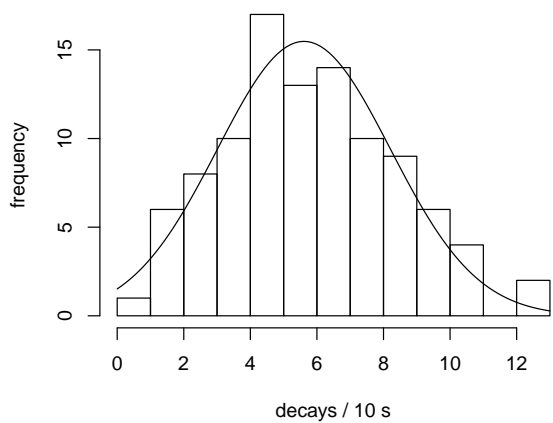}
  \caption{Experimentally obtained histogram of radioactive decay and
    a normal distribution ($\overline{x} = 5.61$, $\sigma_x = 2.60$)
    plotted on top of it.}
  \label{fig:radioaktiv}
\end{figure}

The script shown in Appendix \ref{sec:appc} demonstrates computing and
plotting the histogram in R.  The final result is shown in
figure~\ref{fig:radioaktiv}.  Again, a few comments.
\begin{itemize}
\item Since the format of the input data file is very simple (one
  figure---the number of detected decays per 10~s---per line), we
  have used the \texttt{scan()} function instead of
  \texttt{read.table()}; \texttt{x} is thus a vector rather than a
  data frame.
\item The histogram is plotted using the \texttt{hist()} function.
  The \texttt{breaks} option is used to specify bin size and
  boundaries.  By default, R uses Sturges' formula for distributing
  $n$ samples into $k$ bins:
  \[ k = \lceil \log_2 n + 1 \rceil \; . \]
\item The normal probability distribution function \texttt{dnorm()} is
  plotted over the histogram with the \texttt{curve()} function and
  \texttt{add = TRUE} option.
\end{itemize}

With a direct support for histograms, R is clearly the preferred
solution over a spreadsheet in this case, a fact which would become
even more apparent if the nature of the problem would require
histograms with unequal bin widths.

\section{Discussion}

The amount of information a working physicist---or most any
scientist---has to cope with often exceeds the human capacity of
processing in a tabular form and requires some visualization
technique.  A certain degree of graphic, or visual literacy is
therefore required.  As a result, a fair amount of emphasis is given
to teaching this skill in the course of a physics curriculum
\cite{McDermott:1987,Beichner:1994,KarizMerhar:2009}.  Producing
graphs manually involves some processing of data with a pocket
calculator, determining the minimal and the maximal values of data in
$x$- and $y$ direction, choosing the appropriate scales on the $x$-
and $y$-axis, plotting the data points, drawing the best-fit line and
possibly using it in the subsequent analysis.  In the case of
histogram plotting, the student has to choose the bin size and manually
arrange the samples into bins.  It is instructive to perform this
whole process once or possibly a few times in order to get familiar
with all the steps required.  However, we can not overlook the fact
that the processes involved in a manual production of a graph are
time-consuming, error-prone, and do not contribute significantly to a
better understanding of the problem studied.  In this respect, we have
to agree with earlier studies \cite{Barton:1998} that if the purpose
of the graph is data analysis rather than acquiring the skill of
plotting a graph, the time spent with manual graph plotting is better
spent discussing its meaning.

By employing the ``variable equals spreadsheet cell'' para\-digm where
users can see their variables and their contents on screen,
spreadsheets have been extremely successful in bringing data
processing power to the profile of users who would have otherwise not
embraced a traditional approach to computer programming.  The
usefulness of spreadsheets in education in general (\emph{cf.}
\cite{Baker:2003} and the references therein) and for teaching physics
in particular \cite{Dory:1988,Webb:1993,Cooke:1997}, including
laboratory courses \cite{Guglielmino:1989,Krieger:1990}, has been
recognized early on.

However, the approach taken by spreadsheets also has its drawbacks.
By placing emphasis on visualizing the data itself, the relations
between data are less emphasized than in traditional computer
programs, and it is generally more difficult to debug a spreadsheet
than a traditional computer program of comparable complexity
(\emph{cf.} \cite{Powell:2008a} and the references therein).  The fact
that the variables are referred to by their grid address rather than
by some more meaningful name is an additional hindrance factor.
Leaving aside the questioned validity of some of statistics functions
built into Excel
\cite{Knusel:2005,McCullough:2002,McCullough:2008,Yalta:2008}, which
is outside the scope of this paper, we wish to deal with another sort
of criticism concerning the graphical output of spreadsheet programs.
While a skilled user can produce clear, precise and efficient graphs
with Excel (\emph{cf.} \cite{Vidmar:2007} and the references therein),
it is perceived \cite{Su:2008} that less skilled users are easily
misled into excessive use of various embellishments (dubbed as
``chartjunk'', \cite{Tufte:2001}) that do not add to the information
content.  Telling good graphs from bad ones is not merely a subject of
aesthetics, there exists an extensive body of work in this area
(\emph{cf.}
\cite{Tukey:1977,Chambers:1983,Cleveland:1993,Cleveland:1994,Tufte:2001,Wilkinson:2005}).

Where does R stand in comparison?  As expected from a statistics
suite, R outperforms Excel in terms of statistical accuracy
\cite{Keeling:2007}.  R also does a much better job at graphing.
Except for the obvious required modifications like axes labels, the
default settings are often already satisfactory.  We have nevertheless
demonstrated in figure~\ref{fig:electrolyte} how features like grid
lines can be added, should they be considered necessary.  Excel
charts, on the other hand, come by default with an obtrusive grey
background and horizontal grid lines (as seen in
figure~\ref{fig:histogram}), which do not convey any information
(``non-data ink'', in Edward Tufte's terminology).  In terms of
ease-of-use, Excel excels if the relationship between the data assumes
any of the four pre-defined possibilities: linear, polynomial,
exponential or logarithmic.  Aside from these, more or less clever
techniques need to be adopted in order to use Excel for data analysis,
most of them being too complex for the students to be expected to
invent them on their own.  In R, even the linear modelling requires a
few lines of code.  On the other hand, however, non-linear modelling
is hardly any more complex than linear modelling for the end user, and
is considerably easier than in Excel.  The same is also true for
plotting histograms.

The potential of R in a classroom environment has already been
examined in various disciplines, \emph{e.g.,} statistics
\cite{Horton:2004}, econometrics \cite{Racine:2002}, and
computational biology \cite{Eglen:2009}.  The extent of its use in
student laboratories throughout the world is, however, hard to
estimate.  On the other hand, we can observe that R receives an
explosive growth of use in research laboratories, where one would
actually expect slower adoption due to the additional competition it
faces both from the commercial scientific graphing software and from
the commercial statistical software suites.  Thomson Reuters ISI Web
of Science\textsuperscript{\textregistered} shows that the R manual
\cite{R:Manual} has accumulated over 10.000 citations since 1999,
over 4000 of these in the year 2009 alone.

The main obstacle preventing its more widespread adoption is the fact
that students entering the laboratory course usually have no previous
experience with R, while they usually do have previous experience with
spreadsheets.  Even though spreadsheet users essentially engage in
functional programming \cite{PeytonJones:2003}, they generally don't
perceive it as programming.  Experience shows \cite{Eglen:2009} that
the students can not be expected to self-learn R even in the case of
graduate students of computational biology and that a quick
introduction course is recommended.  The best synergy might be
achieved when students take an introductory course in statistics
before the physics laboratory course or simultaneously with it.

Before we conclude, we would like to mention another weak aspect of
the spreadsheet paradigm: collaboration in spreadsheet authoring is
inherently difficult.  The changes are introduced to the spreadsheet
on the level of spreadsheet cells, and a minor change, replicated
across hundreds or thousands of cells, renders traditional revision
control systems practically useless, meaning it very difficult to
determine who changed what and when did the change occur.  R scripts
are ASCII text, and revision control systems either as basic as RCS
\cite{Tichy:1985} or arbitrarily more complex (CVS, Subversion,
BitKeeper, \emph{etc.}) can be easily employed.  By running scripts in
batch mode rather than using R interactively, reproducible research
techniques \cite{Schwab:2000} are encouraged.  One further step in
this direction is Sweave \cite{Leisch:2003}, which allows creating
integrated script/text documents.

\section{Conclusions}

In the paper, two methods of data analysis and graphing were compared:
spreadsheet software, represented by Microsoft Excel, and a statistics
software suite, represented by GNU R.  Both methods were tried on
three typical tasks taken from an introductory physics laboratory;
similar tasks can be, however, encountered in any science laboratory
course.  The tasks encompass linear dependence, non-linear dependence
and histogram.  It has been determined that spreadsheet offers an
easier approach when the data model is simple (in our case, linear
dependence).  Beyond a few simple models, R offers a more
user-friendly solution.  Histogram plotting, considered one of the
basic tools for data analysis, is also inadequately supported in Excel
(the same is true also for other popular spreadsheet software like
OpenOffice Calc and Apple Numbers).  Using default settings, R
produces charts with a higher ratio of data conveying information to
data conveying no information.

Spreadsheets are often the only software solution for data analysis
students encounter during their education.  We believe the described
shortcomings of spreadsheets make it worth exposing them to other
existing solutions as well.  A statistical software suite is a
solution which can most easily replace a spreadsheet in data analysis
and graphing tasks, and in particular their use should be encouraged
in situations where they can be reused in a statistics course.  Among
the statistical software, we emphasized R.  It is free; students can
install a copy at home.  Using R, well-designed publication-quality
plots can be produced with ease.  It is an object-oriented matrix
language, which encourages thinking on a more abstract level.  The
extent of utilization that R has recently experienced in various
scientific disciplines signifies it is more than a marginal
phenomenon, making it more likely that students who acquire a certain
degree of proficiency with R/S/S-Plus in the course of their studies
will be able to make use of this skill later in their careers.

\section*{Supplementary material}

Worked-out examples employing spreadsheet are available free of charge
on the IOP web site.  An Excel (version 11; Microsoft Office 2003)
spreadsheet \texttt{supplement1.xls} contains four worksheets with the
solutions for linear regression, two solutions for a two-expo\-nential
functions (``2exp log'' and ``2exp nonlin'') and a histogram.


\section*{References}
\bibliographystyle{unsrt}
\bibliography{graphing}

\appendix

\section{Linear dependence}
\label{sec:appa}

{\small
\begin{listing}{1}
concentration <- c(1.e-3, 2.5e-3, 5.e-3, 7.5e-3, 1.e-2)
resistance <- c(5.31, 2.66, 1.40, 0.92, 0.78)

# calculate the conductivity
k <- 1.e-4*resistance[1]
conductivity <- k/resistance

# define the labels on the x- and y-axis
xlabel <- expression(paste(italic(c)," [mol/L]"))
ylabel <- expression(paste(sigma," [(",Omega," cm)"^-1,"]"))

# plot the data
plot(concentration, conductivity,
     xlab = xlabel, ylab = ylabel, xaxs = "i", yaxs = "i",
     xlim = c(0, 0.012), ylim = c(0, 8.e-4))
grid(col = "darkgray")

# calculate the best-fit line
cond.fit <- lm(conductivity ~ concentration,
      data = data.frame(concentration, conductivity))

# plot the best-fit line
abline(cond.fit)
\end{listing}
}

\section{Nonlinear dependence}
\label{sec:appb}

{\small
\begin{listing}{1}
C1 <- read.table("capacitor.csv", dec=",", sep=";", header = TRUE)

U <- C1$U..V.
t <- C1$t..s.

U.nls <- nls(U ~ A*exp(-t/t1) + B*exp(-t/t2),
     start = list(A = 5, B = 5, t1 = 10, t2 = 100))

plot(C1, xlab = expression(paste(italic(t)," [s]")),
     ylab = expression(paste(italic(U)," [V]")), xlim = c(0,600))

lines(0:600, predict(U.nls, list(t = 0:600)))
\end{listing}
}

\section{Histogram}
\label{sec:appc}

{\small
\begin{listing}{1}
# detected decays per 10 s
x <- scan("decay.txt")

# subtract the base (normalized per 10 s)
x <- x - 2.68

avg <- mean(x)
stdev <- sd(x)

# draw the histogram
hist(x, breaks = seq(floor(min(x)), ceiling(max(x))),
     xlab = "decays / 10 s", ylab = "frequency")

# overlay the probability density function
curve(length(x)*dnorm(x, mean = avg, sd = stdev), add = TRUE)
\end{listing}
}

\end{document}